\documentstyle[preprint,aps,epsfig]{revtex}
\begin{document}

\tightenlines


\draft

\title{Semi-classical spectrum of integrable systems in a magnetic field. }
\author{D.~Spehner\cite{spehner}, R.~Narevich   
and E.~Akkermans\\ Department of Physics, 
Technion, 32000 Haifa, Israel}

\maketitle

\begin{abstract}
The quantum dynamics of an electron in a uniform magnetic field is studied for 
geometries corresponding to integrable cases. We obtain the uniform
asymptotic approximation of the 
WKB energies 
and wavefunctions for the semi-infinite plane and the disc. These analytical 
solutions 
are shown to be in excellent agreement with the numerical results obtained 
from the Schr\"{o}dinger equations even for the lowest energy states.  
The classically exact notions of bulk and edge states are followed to
their semi-classical limit, when the uniform approximation provides
the connection between bulk and edge. 
\end{abstract}


\def\real{{\rm I\kern-.2em R}}
\def\complex{\kern.1em{\raise.47ex\hbox{
	    $\scriptscriptstyle |$}}\kern-.40em{\rm C}}
\def\integer{{\rm Z\kern-.32em Z}}
\def\pinteger{{\rm I\kern-.15em N}}
\newcommand{\arccosh}{\rm{arccosh}}
\newcommand{\arcsinh}{\rm{arcsinh}}
\newcommand{\Ai}{\rm{Ai}}
\newcommand{\Bi}{\rm{Bi}}
\newcommand{\ta}{\eta}
\newcommand{\Tr}{\rm{Tr}}
\newcommand{\im}{\rm{Im}}
\newcommand{\re}{\rm{Re}}

\narrowtext

\section{Introduction.}

The aim of this work is to present some analytical methods to obtain the 
energy spectrum and the eigenfunctions of non-interacting electrons constrained 
to a finite domain with boundaries and submitted to a uniform magnetic field.

This problem is relevant to various situations in 
condensed matter physics. 
In the low magnetic field regime, defined by the condition $\Phi \ll 
{\Phi_0}$, where $\Phi$ is the magnetic flux through the system and 
$\Phi_0 = {hc \over e}$ is the quantum flux, recent 
experiments performed on small metallic systems did show how 
important is the effect of the boundaries \cite{experiments}. It determines the 
nature of the zero-field 
classical motion being either integrable or chaotic. The magnetic susceptibility 
has been shown using numerical and semiclassical methods to be reduced and to present large
fluctuations in the chaotic case, whereas it is larger and varies slowly with the field
in the integrable case \cite{nakamura,jalabert}. 
In the opposite limit $\Phi \gg \Phi_0$ of 
high fields, we are in the so-called integer quantum Hall effect 
regime (IQHE), where the edge states associated with the 
boundary play a prominent role \cite{QHE}. 
In this work, we shall concentrate on the problem of non-interacting 
electrons in a semi-infinite plane and in a disc.

The classical dynamics allows for a natural distinction
between bulk and edge states. A first semiclassical method is based on the
Einstein-Brillouin-Keller (EBK) quantization rules \cite{EBK} and preserves 
this bulk and edge states splitting, by giving different quantization rules 
for each of them. This approximation is further improved for the semi-infinite 
plane by constructing the asymptotically
matched WKB function and then finding its zeros corresponding to the
energy levels. This procedure
removes the gap between the bulk and edge energies, resulting in a 
very good approximation for the exact spectrum. The calculation is not new
\cite{isihara}, however we present it in detail, as it serves a starting
point for the WKB approximation for the disc.
The spectrum of the disc is found using the
more general comparison equation method (for a general reference, see 
\cite{berry mount}), also called the Miller-Good method \cite{miller}
 to map the problem 
onto the semi-infinite plane's one. The obtained semiclassical formulas are 
valid for any strength of the magnetic field, thus providing us a way to 
study more
particularly both the IQHE and the low field regimes.

\section{The WKB spectrum of the semi-infinite plane}

In this section the spectrum of an electron in the semi-infinite plane in a magnetic
field is approximated first using the EBK quantization rules and then
building the
matched WKB wavefunction. To introduce semiclassical language, we begin by 
considering the classical dynamics.

\subsection{The classical dynamics.}
\begin{figure}
\begin{center}
  	\leavevmode
         \epsfig{file=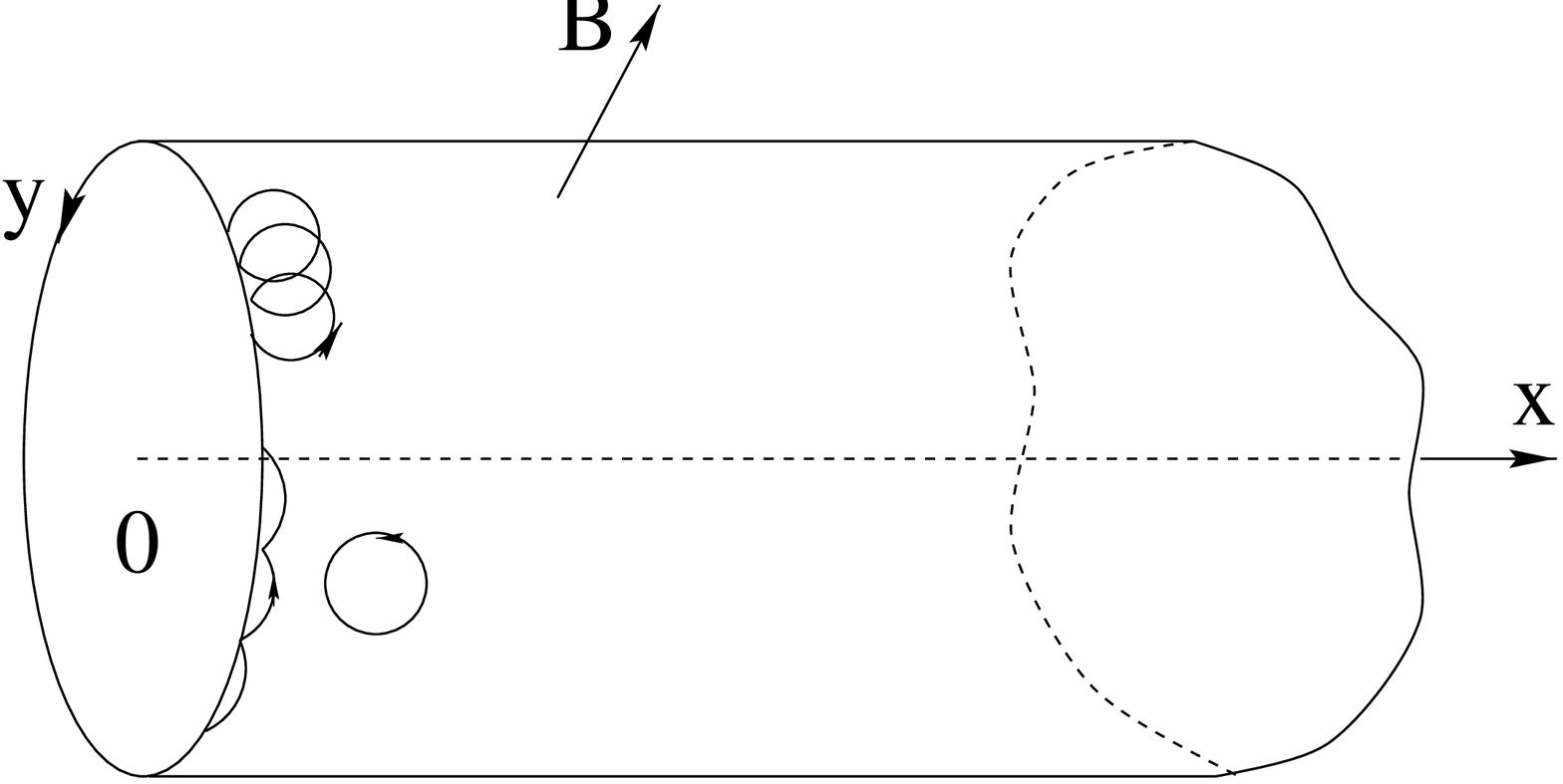,height=10cm,width=5cm,angle=270}
\caption{Semi-infinite cylinder in the magnetic field $B$.}
\label{fig:1}
\end{center}
\end{figure}

We consider a spinless particle of charge $-e$ ($e>0$) and mass $m$
constrained to move in the semi-infinite plane. A uniform magnetic field 
$B$ is applied perpendicular to the plane. Cartesian coordinates are 
defined
such that the $x$ axis is perpendicular to the boundary and the motion is
confined to positive values of $x$. It is convenient to consider the  boundary 
having a finite length $L$ and therefore we impose periodic 
boundary conditions in 
the $y$ direction so that the particle moves on a semi-infinite 
cylinder (Fig. \ref{fig:1}).

In the Landau gauge ${\bf{A}}=(0\,,Bx)$, the 
Hamiltonian of the particle is:
\begin{equation} \label{classical hamiltonian}
H =\frac{1}{2m}\,\left( {{p}_{x}}^{2}+({p}_{y}+\frac{e}{c}Bx)^{2}
\right), \end{equation}
and the momentum is ${\bf{p}}=(m\dot{x}\,,m\dot{y}-\frac{e}{c}Bx)$. 
The total energy $E$ and the $y$-component ${p}_{y}$ of the momentum 
are constants of motion, therefore the problem is integrable. In the 
four-dimensional
phase space of the Cartesian coordinates and the corresponding momenta, 
each family of classical trajectories are winding on an invariant torus
defined by the two constants of motion.


The ensemble of trajectories splits naturally into two families: those 
that do not touch
the boundary (bulk trajectories), and others (edge trajectories).
Bulk particles of energy $E$ go counterclockwise in circles of radius 
$r_{c}=\sqrt{{2E}/{m\omega^2}}\,$ (where $\omega={eB}/{mc}$ is the cyclotron 
frequency) with their center farther than $r_c$ from the 
edge and momenta $p_y < -\sqrt{2mE}$. Edge particles
have momenta $p_y > -\sqrt{2mE}$ and 
undergo  specular reflections on the boundary before 
 closing a circle so that
the cyclotron orbit center begins drifting 
along the edge (Fig. \ref{fig:1}). Increasing $p_y$ at 
fixed energy, the bulk tori in phase space are  transformed at 
$p_y= -\sqrt{2mE}$ in a discontinuous way into edge tori.

\subsection{The EBK quantization.}

We consider now the quantum-mechanical version of the same problem.
Dirichlet boundary conditions are imposed on the wavefunction $\Psi (x,y)$: 
\begin{equation} \label{dirichlet}
\Psi(0,y)=0. \end{equation}  

 The application of the EBK quantization rules for 
integrable systems leads to the quantization of the two actions $I_{y}$ and $I_{x}$
along the $y$ and $x$ axis. The action $I_{y}$ corresponding to the motion 
along the $y$ axis (parallel to the boundary) is
\begin{equation} \label{quantization y}
I_{y} = \frac{1}{2\pi} \oint p_{y}\;dy = \frac{p_{y}L}{2\pi} = n_{y}\hbar ,
\end{equation}  
with $n_{y} \in \integer$. The motion along the $x$ axis 
(perpendicular to the boundary) is different 
for the bulk and the edge states, therefore their quantization differs
too.  In particular the energy of the bulk states is found from
\begin{eqnarray} \label{quantization bulk}
I_{x} &=& \frac{1}{2\pi} \oint p_{x}\;dx  = 
 \frac{1}{\pi} \intop_{0}^{2r_{c}} 
\sqrt{2mE-(p_{y} +\frac{e}{c}Bx)^2}\,dx  = 
\frac{E}{\omega}  =  
\hbar(n_{x} + \frac{1}{2}), 
\end{eqnarray} 
where $n_x \in \pinteger$. For the edge states the EBK condition is 
\begin{eqnarray} 
\label{quantization edge}
I_{x}  & = & \frac{1}{\pi} \intop_{0}^{r_{c}-\frac{p_{y}}{m\omega}} \nonumber
\sqrt{2mE-(p_{y}+\frac{e}{c}Bx)^2}\,dx  = 
\frac{E}{\pi \omega} \left( \arccos \xi_{0} - \xi_{0} 
\sqrt{1-\xi_{0}^{2}} \right)  \\ &= &
\hbar(n_{x} + \frac{3}{4}), 
\end{eqnarray}
where 
\begin{equation}
\label{xi0}
\xi_{0} = \frac{p_{y}}{m\omega r_{c}} = \frac{2\pi n_{y} L}{\sqrt{2mE}}.
\end{equation}
For the bulk trajectories ($\xi_0 <-1$), the Maslov index is $\frac{1}{2}$ and 
we obtain degenerate Landau levels 
$E = \hbar \omega (n_{x} + \frac{1}{2})$ 
which correspond to states that do not feel the boundary.  
For the edge trajectories ($|\xi_0| < 1$), the integration range is restricted because of the 
reflection on the boundary and
the Maslov index is $\frac{3}{4}$, since
there is one turning point (Maslov index $\frac{1}{4}$)  and one reflection (Maslov 
index $\frac{1}{2}$ associated to a change of sign of the wavefunction).
The energies are  
implicit solutions of (\ref{quantization edge}), they are non degenerate and bounded
below by $\hbar \omega (n_{x} + \frac{3}{4} )$ for each $n_{x}$.
Note that $E \rightarrow \infty$ when $\xi_0 \rightarrow 1$. 
There is a singularity in the EBK spectrum, separating bulk 
and edge energies.

\subsection{Matching the WKB wavefunctions.}

Once the motion in the $y$-direction is integrated, the Schr\"{o}dinger 
equation  together with the boundary condition (\ref
{dirichlet}) reduces to a one-dimensional Sturm-Liouville problem. A 
systematic
WKB analysis is well-developed for those kind of problems 
and improves the EBK quantization. 

We introduce the dimensionless variable 
$\tilde{x}={\sqrt{2} \,x}/{l_B}$, and set $\tilde{x_0}= {\sqrt{2} \,p_{y}}/
{m\omega l_B}$ and 
$\epsilon={\hbar\omega}/{2E}$, where $l_B=\sqrt{{\hbar c}/{eB}}$ is the magnetic 
length. Using (\ref{classical hamiltonian}), the
 Schr\"{o}dinger equation 
for the one-dimensional wavefunction $\varphi(\tilde{x}) = 
e^{-{ip_{y}\,y}/{\hbar}}\,\Psi(x,y) $ reads:
\begin{equation} \label{transformed equation} 
\left( \frac{d^{2}}{d\tilde{x}^{2}} + \frac{E}{\hbar\omega} - 
\frac{1}{4}\,(\tilde{x}+\tilde{x}_0)^{2} \right) \,  \varphi(\tilde{x}) =0. 
\end{equation}
A subsequent change of variable $\xi = 
\sqrt{\frac{\epsilon}{2}}\,(\tilde{x}+\tilde{x}_0) - 1 = \frac{x}{r_c}+ 
\xi_0 -1$ gives:
\begin{equation} \label{wkb equation}
\epsilon^2\, f''(\xi) - (\xi^2+2\xi)\,f(\xi) = 0,  
\end{equation}
where $f(\xi)=\varphi(\tilde{x})$. Eq. (\ref{transformed equation}) is a
Weber equation \cite{bender,abramowitz}, and its solutions 
(vanishing at infinity) are: $\varphi(\tilde{x}) = c \,D_{\frac{E}{\hbar\omega}-\frac{1}{2}} 
(\tilde{x}+\tilde{x}_0)$, where 
$c$ is a constant and $D_{\nu}(u)$ a parabolic cylinder function. 
The Dirichlet boundary condition (\ref{dirichlet}) reads in the new variables:
\begin{equation} \label{dir1} 
\varphi(0)= f(\xi_{0}-1)  = 0.
\end{equation} 
Since the solutions $f$ of 
Eq. (\ref{wkb equation}) depend 
only on $\epsilon$, without this condition the energies would be 
proportional to 
$\hbar \omega$ (Landau levels). However this condition makes 
the rescaled energies $\frac{E}{\hbar\omega}$ to depend on the 
(energy-dependent) parameter $\xi_0$.
They are therefore given by an implicit equation of the form: 
\begin{equation} \label{energy band}
\frac{E}{\hbar \omega} = \frac{1}{2 \epsilon} = h_{n_x}(\xi_0),
\end{equation} 
where $\xi_0$ is given by 
 (\ref{xi0}), and the label $n_x$  refers to
different `energy bands' (in the infinite $L$ limit, 
$h_{n_x}(\xi_0)$ gives the appropriate energy bands by continuously 
varying $\xi_{0}$). 

Our purpose in this section is to use a semiclassical approximation in 
order to find 
explicit analytical 
expressions for the energies. As noted by Isihara and Ebina \cite{isihara},
for small $\epsilon$ (i.e., large energies), (\ref{wkb equation}) is a standard example of 
equation where the WKB method is applicable.  It has two turning points  at $\xi_1=0$ and 
$\xi_2=-2$. Sufficiently far from them, the WKB function is given by the 
following asymptotic expression
\cite{bender}:
\begin{equation}   \label{general wkb function}
f_{WKB}(\xi)=(\xi^2+2\xi)^{-{\frac{1}{4}}} \, \left( c_1 \, 
e^{-\frac{1}{\epsilon} \, 
\intop^{\xi} dt \,\sqrt{t^2+2t}}  +c_2 \, e^{\frac{1}{\epsilon} \,
\intop^{\xi} d t\,\sqrt{t^2+2t}} \, \right),
\end{equation}
where $c_{1}$ and $c_{2}$ are arbitrary constants. 
In the vicinity of the turning points this approximation breaks down.
The potential $(\xi^2+2\xi)$ is then linearized in Eq. (\ref{wkb equation}) 
and the resulting Airy equation can be solved exactly. 
One obtains the approximate solution $f_{WKB}(\xi)$ by matching the corresponding Airy 
solutions for each turning point with the WKB solutions
(\ref{general wkb function}) valid to the right, between, and to the left of 
them. 

The function $f_{WKB}(\xi)$ consists of five branches, related to 
the five overlapping
intervals in which the WKB and Airy approximations are valid. 
We start from large $\xi$'s.
\begin{itemize}
\item[i.]
The first branch has a domain defined by $\xi \gg \epsilon^
{\frac{2}{3}}$, where the expression (\ref{general wkb function}) holds 
\cite{bender}.
The vanishing of the wavefunction at infinity fixes $c_2=0$, and we set
arbitrarily $c_1=1$. Thus, the first branch of the WKB function is: 
\begin{equation}  \label{wkb1}
f_{WKB}^{(1)}(\xi)=(\xi^2+2\xi)^{-{\frac{1}{4}}} \, 
 e^{-\frac{1}{\epsilon} \,
\intop^{\xi}_{0} dt \,\sqrt{t^2+2t}}.  
\end{equation}
It is always nonzero, so there are no solutions of Eq. (\ref{dir1}) 
for $\xi_{0} -1 \gg \epsilon^{\frac{2}{3}}$.

\item[ii.]
The second branch approximates the exact 
solution near the first turning point, i.e. for  $|\xi| \ll 1$, where
$\xi^2+2\xi$ can be linearized and replaced by $2\xi$ in equation (\ref{wkb equation}). 
It consists therefore of a linear combination of two independent
Airy functions Ai($t$) and Bi($t$) with $t= \frac{ 2^{{1}/{3}} }
{ \epsilon^{{2}/{3}} } \,\xi $.
When $\xi$ is such that $ \epsilon^{\frac{2}{3}} \ll \xi \ll 
\epsilon^{\frac{2}{5}}$, $t$  is large so the asymptotic approximations of Airy functions 
can be used, and we can approximate the exponential in (\ref{wkb1}) by
$\exp (-\frac{2 \sqrt{2}}{3 \epsilon}\, \xi^{\frac{3}{2}} )$. We 
match for such $\xi$'s the solution of the linearized equation with $f_{WKB}^{(1)}
(\xi)$ and determine the unknown constants of the linear combination. 
This gives:
\begin{equation}  \label{wkb2}
f_{WKB}^{(2)}(\xi)=\frac{2\sqrt{\pi}}{(2\epsilon)^{\frac{1}{6}}} \, 
\Ai (\frac{2^\frac{1}{3}}{\epsilon^{\frac{2}{3}}}\xi).
\end{equation}
The energies are found by imposing the Dirichlet condition on this solution:
\begin{equation}  \label{wkb2 energies}
\frac{E}{\hbar \omega} = (\frac{a_{n_x}}{2\xi_{0}-2})^{\frac{3}{2}},\,\,\, 0 < 1-\xi_{0} 
\ll 1,
\end{equation}
where the $a_{n_x}$ are the $n_x$'th zeros of the Airy function Ai($t$). 
These energies are associated
with the `whispering gallery' orbits (see Fig. \ref{fig:1}) which are 
concentrated in a very narrow region near the boundary.
We notice that the energies (\ref{wkb2 energies})
can be obtained from a `generalized EBK rule'  by 
letting $n_x$ be such that $n_x + \frac{3}{4} = 
\frac{2}{3\pi}(-a_{n_x})^{\frac{3}{2}}$ in 
(\ref{quantization edge}), instead of an integer. Note that there are no energies 
for $\xi_{0} > 1$, since the real zeros of Ai($t$) are negative.

\item[iii.]

Between the two turning points $\xi_1$ and $\xi_2$, the expression (\ref
{general wkb function}) holds again and to determine the constants $c_1$ and $c_2$
we do the matching with $f_{WKB}^{(2)}(\xi)$ in the interval
$-\epsilon^{\frac{2}{5}} \ll \xi \ll -\epsilon^{\frac{2}{3}}$. This gives for
$\xi \ll -\epsilon^{\frac{2}{3}}$ and $\xi + 2 \gg \epsilon^{\frac{2}{3}}$ :  
\begin{equation} \label{wkb3}
f_{WKB}^{(3)}(\xi) = 2 (-\xi^2-2\xi)^{-{\frac{1}{4}}} \,  \sin({\frac{1}
{\epsilon} \, \intop^{0}_{\xi} dt \,\sqrt{-t^2-2t} + \frac{\pi}{4}}).  
\end{equation}
The argument of the sine for $\xi = \xi_0 -1$ is 
$\pi\,(\,\frac{I_x}{\hbar} + \frac{1}{4}\,)$, where
$I_x$ is the classical action for the edge motion calculated in (\ref{quantization edge}). 
Thus as expected we obtain the EBK quantization 
(\ref{quantization edge}) for $\xi_0$ between but not too close to $1$ and $-1$.
The comparison of the energies derived within
the WKB approximation with the exact ones found numerically is shown in 
Fig. \ref{fig:2a}.



\item[iv.]

This branch represents the function in the vicinity of the second turning point
$\xi_2=-2$. Repeating the scheme of matching with the previous third branch, we 
obtain for $|\xi+2| \ll 1$:
\begin{eqnarray}  \label{wkb4}
f_{WKB}^{(4)}(\xi)=\frac{2\sqrt{\pi}}{(2\epsilon)^{\frac{1}{6}}} \, \left( \sin
(\frac{\pi}{2\epsilon})  \Ai (-\frac{2^\frac{1}{3}}{\epsilon^{\frac{2}{3}}}(\xi+
2)) 
+\cos (\frac{\pi}{2\epsilon}) \Bi (-\frac{2^\frac{1}{3}}
{\epsilon^{\frac{2}{3}}}(\xi+2))  \right).  
\end{eqnarray} 
The energies correspond here either to edge trajectories which nearly 
complete full circles before being reflected or to bulk trajectories 
very close to the boundary (fig. \ref{fig:1}).
In Fig. \ref{fig:2b} we see that, for the lowest band ($\epsilon \sim  1$),
the energies of the fourth 
branch, found by setting $f_{WKB}^{(4)}(\xi_0-1)=0$, become closer below $\xi_0 \simeq -0.5$
to the exact energies than those of the third branch.


\item[v.]

The fifth branch $f_{WKB}^{(5)}(\xi)$ is derived in a similar way. 
For $-\xi -2  \gg \epsilon^{\frac{2}{3}} $:
\begin{eqnarray}  \label{wkb5}
f_{WKB}^{(5)}(\xi)= ( \xi^2+2\xi)^{-{\frac{1}{4}}} \, \left( \sin
(\frac{\pi}{2\epsilon}) \, e^{-\frac{1}{\epsilon} \, \intop_{\xi}^{-2} dt \,
\sqrt{t^2+2t}}   + 2 \cos
(\frac{\pi}{2\epsilon}) \, e^{\frac{1}{\epsilon} \, \intop_{\xi}^{-2} dt \,
\sqrt{t^2+2t}} \right).  
\end{eqnarray}
The equation to be solved for the energies is:
\begin{equation}  \label{wkb5 energies}
\frac{1}{2} \tan \frac{\pi}{2\epsilon} = -e^{\frac{1}{\epsilon}(-\arccosh 
(-\xi_0) - \xi_0 \, \sqrt{\xi_{0}^{2}-1} )}, \,\,\, -\xi_0-1 \gg \epsilon^{\frac{2}{3}}.   
\end{equation}
When $|\xi_0|$ is large, the right-hand side of this equation is a 
large number so that the argument of the tangent in the left-hand side is to
a good approximation $\pi$ time an half-integer. Thus for $|\xi_0|$ sufficiently large
the energies are very close to the Landau levels given by 
(\ref{quantization bulk}), i.e. $1/{2\epsilon}=n_x+1/2$. 
The solutions of the energy equations for the fourth and the fifth branches and
the exact energies are shown in Fig. \ref{fig:2c}.


\end{itemize}

The graphs in Fig. \ref{fig:2a}, \ref{fig:2b} and \ref{fig:2c} 
show a very good agreement between the exact 
calculation and the matched WKB approximation even for the lowest energies.
The distinction between the bulk and the edge states
disappears and the spectrum rises smoothly from $\frac{\hbar\omega}{2}$ (from 
above) to infinitely large edge energies.

\section{Semiclassical spectrum of the disc in magnetic field}

We study here the similar problem as that of the previous section but for a disc of 
radius $R$ and a magnetic field applied in the $z$-direction perpendicular to the disc. 
For this geometry, a convenient choice for ${\bf{A}}$
is given by the radial gauge ${\bf{A}} = \frac{1}{2}Br \,{\bf{e}}_{\theta}$, where $(r,\theta)$
are polar coordinates and ${\bf{e}}_r$, ${\bf{e}}_{\theta}$ the corresponding
unit vectors. The two momenta canonically conjugated to $r$ and $\theta$ are 
$p_r = m \dot{r}$ and 
$p_{\theta}= m r^2 (\dot{\theta}-\frac{\omega}{2} )$, the $z$-component of the
 angular 
momentum. In the symmetric gauge the Hamiltonian of 
the particle reads:
\begin{equation} \label{dhamiltonian}
  H=\frac{1}{2m}\,\left( {{p}_{r}}^{2}+\frac{1}{r^2}({p}_{\theta}+
 \frac{m \omega}{2} r^2 )^{2} \right).
\end{equation}
Since $p_{\theta}$ and the energy are conserved by reflexion on the boundary, we again 
deal with an integrable system.

We can as before distinguish between edge and bulk trajectories:
 for $-m\omega(\frac{1}{2}R^2+R r_c) < p_{\theta}
< -m\omega (\frac{1}{2}R^2-R r_c)$ the particle bounces
and drifts along the boundary (edge trajectories); for 
$-m\omega (\frac{1}{2}R^2-R r_c) < p_{\theta} \leq \frac{E}{\omega}$ it follows a closed-circle 
bulk trajectory; for $p_{\theta} > \frac{E}{\omega}$ or
$p_{\theta} < -m\omega(\frac{1}{2}R^2+R r_c)$, the motion is forbidden or outside the disc. 
Note that positive values of angular momenta are associated with bulk trajectories 
whose cyclotron orbit centers are closer than $2r_c$ to the origin. Let us emphasize that if 
$R$ is of the order of or less than the magnetic length $l_B$, the distinction between edge and 
bulk motion becomes meaningless from the quantum-mechanical point of view, all 
cyclotron radii being greater than $l_B$, and then also than $R$ 
(${\hbar \omega}/2$ is a lower bound for the energy). For such low fields or 
small systems 
all classical trajectories of interest are of the edge type. The case of weak
and zero fields has been extensively studied before \cite{jalabert,keller}.
We will return to the weak field limit later to show the limitation
of our methods in that regime.

We now formulate the quantum-mechanical problem. Since the system is invariant
 by rotation along 
($Oz$), we choose wavefunctions $\Psi(r,\theta)= \varphi(r) e^{il\theta}$.
We introduce the dimensionless variables $\xi=\frac{r}{r_c}$, 
$\epsilon=\frac{\hbar\omega}
{2E}$ and $l_z= \frac{l \hbar}{m \omega r_c^2} = l \epsilon$. 
The wavefunction $\varphi(\xi)$ then satisfies
the Schr\"{o}dinger radial equation:
\begin{equation} \label{dsturm liouville}
\left(
\epsilon^2 ( \frac{d^2}{d\xi^2}+\frac{1}{\xi}\,\frac{d}{d\xi}\,)-\frac{\xi^2}{4}
+1-l_z - \frac{l_z^2}{\xi^2}\,
\right)
\varphi(\xi) = 0.
\end{equation}
We impose the boundary conditions  $\lim_{\xi \rightarrow 0} \,\varphi(\xi) 
 < \infty$
and  $\varphi(\frac{R}{r_c}) = 0$. 
Note that equation (\ref{dsturm liouville}) with these boundary conditions 
has an exact solution given by
$\varphi(\xi)= c\, {\xi}^{|l|} e^{-\frac{\xi^2}{4\epsilon}}\, _1 F_{1}\,
(\frac{1+l +|l|}{2}-\frac{1}{2\epsilon} , 1 + |l| ; \frac{\xi^2}
{2 \epsilon})$, where $c$ is a constant and $_1F_1\,(a,c;u)$ the confluent
hypergeometric function \cite{abramowitz}. The semiclassical approximation
of Eq. (\ref{dsturm liouville}) becomes better as the magnetic field is
decreased. Indeed, the energies $E$ tend to their finite zero-field values 
when $B \rightarrow 0$ and therefore
$\epsilon=\hbar\omega/E$ approaches zero.

As a result of the singularity at the origin due to 
the centrifugal potential $\frac{l_z^2}{\xi^2}$ in equation 
(\ref{dsturm liouville}), the WKB method
fails when applied directly. 
To overcome this difficulty we follow Langer \cite{langer} and make the change of 
variable  
$x = - \ln(\frac{\xi^2}{a})$, where $a= 2(1-l_z)$. The resulting equation for the function
$f(x)= \varphi(\xi)$ is:
\begin{equation} \label{dwkb equation}
(\frac{4\epsilon}{a})^2 
\frac{d^2 f}{dx^2} = Q(x) f(x), 
\,\,\,\,\,  Q(x) = e^{-2x} - 2 e^{-x} + (\frac{2l_z}{a})^2. 
\end{equation}
This equation can be solved using a WKB analysis similar to that of section 2.
The two turning points are $x_{\pm} = -\ln (1\mp c)$, where $c=\sqrt{1-(\frac{2l_z}{a})^2}$.
Note that in two dimensions (the case we consider here) there is no need for a Langer substitution as in the 
case of the radial equation for the hydrogen atom \cite{kramers}.
For $l_z>\frac{1}{2}$, there are no real turning points, 
thus the WKB function is like in (\ref{wkb1}) an exponentially increasing function when 
$x\rightarrow -\infty$; therefore there are no 
solutions for such $l_z$'s,
and we restrict our discussion to the classically allowed region 
$l_z\leq \frac{1}{2}$.

However we shall prefer a uniform approximation approach to the WKB
analysis, because the two turning 
points $x_+$ and $x_-$ coalesce when $l_z\rightarrow 
\frac{1}{2}$ and $l_z\rightarrow -\infty$ (the value $l_z= \frac{1}{2}$ corresponds to a
cyclotron trajectory centered at the origin; $l_z\rightarrow -\infty$ is actually not 
relevant). $l_z=0$ is also a special case where one of the turning points 
($x_{-}$) 
goes to infinity.
We make use here of the so-called comparison 
equation method, that gives correct approximations no matter how close 
are the turning points. This will enable us to map the problem of the 
disc to the previously solved semi-infinite plane case. 
In this method \cite{berry mount}, the approximate solution is written in 
terms of a function $g(\sigma)$,
solution of a simpler equation. We shall choose for this simpler 
`comparison equation' the
equation (\ref{wkb equation}) we already solved, in which we make for 
convenience the change 
of variable $\bar{\xi} = \sqrt{\bar{\epsilon}} \,\sigma -1$ so that
 $\frac{d^2g}{d\sigma^2} = 
( \sigma^2 - \frac{1}{\bar{\epsilon}}) g(\sigma)$, where $\bar{\epsilon}$ is 
a parameter
to be determined later.
The solution of (\ref{dwkb equation}) is obtained by multiplying  
$g(\sigma)$ by a slowly varying amplitude and allowing its argument $\sigma$ to 
depend weakly on $x$. In other words, we look for solutions of the form
 $f(x)= A(x) g(\sigma(x))$. 
The equation obeyed by the mapping function $\sigma(x)$ 
is simplified by choosing $A(x) = (\frac{d\sigma}{dx})^{-\frac{1}{2}}$ and reads:
\begin{equation} \label{mapping equation}
(\frac{a}{4\epsilon})^2
Q(x)
= (\frac{d\sigma}{dx} )^2 \left( \sigma^2 - \frac{1}{\bar{\epsilon}} \right)
+(\frac{d\sigma}{dx})^{\frac{1}{2}} \frac{d^2}{dx^2} 
(\frac{d\sigma}{dx})^{-\frac{1}{2}}.
\end{equation}
If $(\frac{d\sigma}{dx})^{-\frac{1}{2}}$ varies slowly enough or if the 
left-hand side
is large enough, the second term in the right-hand side of 
equation (\ref{mapping equation}) is negligible. This constitutes the approximation itself 
(for more details see \cite{berry mount}).

We define by analogy with (\ref{quantization edge}) the classical radial action variable:
\begin{eqnarray} \label{daction}
&& \pi I_{r} (x_0) =  \oint p_r \; dr = 
\intop_{r_c\sqrt{a(1 - c)}}^{r_c\sqrt{a} e^{-\frac{1}{2}x_0}} 
p_r \;dr  \nonumber  =  \frac{a m \omega r_c^2}{4} 
\intop_{x_0}^{x_{+}}  \;\sqrt{-Q(x) } \; dx \\
 & & =
\frac{ a m \omega r_c^2}{4} 
\left(
\sqrt{c^2 - (e^{-x_0} -1)^2} + \arccos (\frac{1-e^{-x_0}}{c}) - \sqrt{1-c^2} \, \arccos (
\frac{1-c^2}{c} e^{x_0} -\frac{1}{c})
\right)
\end{eqnarray}
(if $x_0 >x_{+}$, this action is purely imaginary and we use 
$\arccos(u) = -i \,\arccosh(u)$, $\sqrt{1-u^2}= -i \,\sqrt{u^2-1}$ for all 
$u>1$; if $x_0< x_{-}$, we define 
$I_r(x_0)$ by 
integrating from $x_0$ to $x_{-}$ so it is again purely
imaginary and (\ref{daction}) still holds with $\arccos (u) = i\,\arccosh|u|$,
$\sqrt{1-u^2}= -i \,\sqrt{u^2-1}$ for
$u<-1$ ).



For the above analysis to be meaningful, the mapping $x\mapsto \sigma$ has to be 
one-to-one, i.e. $\frac{d\sigma}{dx}$ and $\frac{dx}{d\sigma}$ must be always
non-zero.  From 
(\ref{mapping equation}) it follows that $x_{\pm}$ must map into 
$\pm\frac{1}{\sqrt{\bar{\epsilon}}}$. This enables us to integrate our approximation of 
(\ref{mapping equation}), which takes the form:
\begin{equation} \label{integrated map equation}
I_r(x_0) = \bar{I_x}(\, \bar{\xi_0} = \sqrt{\bar{\epsilon}} \,\sigma(x_0)\, ) =
\frac{\hbar}{2 \pi \bar{\epsilon} } ( \arccos \bar{\xi_0} - \bar{\xi_0} \sqrt{1-\bar{\xi_0}^2}), 
\end{equation}
where $\bar{I_x}(\bar{\xi_0})$ is the classical action 
(\ref{quantization edge}) for the 
semi-infinite plane. The energy $\frac{1}{2\bar{\epsilon}}$ is obtained 
by letting 
$\sqrt{\bar{\epsilon}} \,\sigma(x_{-})= -1$ in (\ref{integrated map equation}). 
Taking 
$\theta(l)=0$ for $l<0$ and $1$ for $l\geq 0$, we obtain
\begin{equation} \label{hatenergy}
\frac{1}{2 \bar{\epsilon}} = \frac{1}{2 \epsilon} - l \,\theta(l).
\end{equation} 
The mapping function takes the following forms for $x \rightarrow
\pm \infty$ and $x \rightarrow x_{\pm}$:
\begin{eqnarray} \label{approximate sigma}
\sqrt{\bar{\epsilon}} \, \sigma(x) & \nonumber \simeq & (\bar{\epsilon} |l| x)^{\frac{1}{2}}
\,\,\,{\rm{if}} \,\, x \gg 1 \,\,{\rm{and}} \,\, x \gg \frac{1-l_z}{|l_z|} \\
\sqrt{\bar{\epsilon}} \, \sigma(x) & \nonumber \simeq & -\sqrt{ \frac{a \bar{\epsilon}}
{2 \epsilon }} \, e^{-\frac{x}{2} } \,\,\, {\rm{if}} \,\, x \ll -1 \\
\sqrt{\bar{\epsilon}} \, \sigma(x) \mp 1 & \simeq & \left( \frac{a^2 c \,\bar{\epsilon}^2}
{16(1 \mp c)^2 \epsilon^2 } \right)^ {\frac{1}{3}} (-e^{-x}+1\mp c) \,\,\,{\rm{if}} \,\,
|e^{-x}-1 \pm c| \ll c.
\end{eqnarray}
The semiclassical wavefunction is then given (up to an arbitrary multiplicative constant) by:
\begin{equation} \label{dwkb}
f_{WKB}(x) = \left| \frac{\bar{\epsilon} \,\sigma^2(x) -1 }{Q(x)} \right| ^{\frac{1}{4}} 
\bar f_{WKB} (\,\bar{\xi} = \sqrt{\bar{\epsilon}} \, \sigma(x) -1\,),
\end{equation}
where $\bar f_{WKB} (\bar{\xi})$ is the WKB solution of equation
(\ref{wkb equation}). Moreover, since $\lim_{x \rightarrow 
\infty} \sigma (x) = +\infty $, it satisfies the same boundary conditions as 
in section 2, i.e.  
$\lim_{\bar{\xi} \rightarrow \infty} \bar f_{WKB} (\bar{\xi})  < \infty$ and
$\,\,\bar f_{WKB} (\bar{\xi_0}-1) = 0$, with:
\begin{equation} \label{hatxi0}
x_0 = -\ln (\frac{R^2}{ar_c^2}) = \ln (\frac{1}{\epsilon N_{\Phi}}-\frac{l}{N_{\Phi}}) ;
\,\,\,\,\, 
\bar{\xi_0} = \sqrt{\bar{\epsilon}} \,\sigma(x_0),
\end{equation}
where $N_{\Phi}=\frac{m\omega R^2}{2 \hbar}$ is the total magnetic flux through the system in units of 
the flux quantum. 

To find the set of discrete levels one has therefore to solve the implicit 
equation
\begin{equation}
\frac{1}{2\epsilon} = h_{\bar{n}_x} (\bar{\xi_0})+l\,\theta(l), 
\end{equation}
where    
$\bar{\xi_0}$ depends on $\epsilon$ through (\ref{hatxi0}). $h_{\bar{n}_x}(\bar{\xi_0})$   
is the band energy curve (\ref{energy band}) for the semi-infinite plane, and also the band
energy curve for the disc 
in the limit of infinitely large $N_\Phi$ if $l_z<0$. In fact, in this limit $\bar{\xi_0}$ 
in (\ref{hatxi0}) varies continuously when $l$ runs over all negative integer values 
($\sigma$ is continuous).
We distinguish the following typical cases:
\begin{itemize}
\item[i.] 
$N_{\Phi} \gg 1$, {\em{bulk states}}. \\ For absolute values of the angular momentum 
$|l| \ll N_{\Phi}$ and energies low enough for $N_{\Phi}\,\epsilon= \frac{R^2}{2 r_c^2}
\gg 1$ to hold,
$x_0$ and $\bar{\xi_0}$ are large negative 
numbers (see (\ref{hatxi0}), (\ref{hatenergy}) and (\ref{approximate sigma})).
 It follows from 
the analysis of the previous section that $h_{n_x}(\bar{\xi}_0)$ 
is close to a Landau level $\bar{n}_x+\frac{1}{2}$, i.e. levels with small 
angular momenta 
$| l | \ll N_{\Phi}$ practically do not feel the boundary. For negative $l$, 
the level is close 
to the $\bar{n}_x$'th Landau level, for positive $l$ the level is close to the 
$(\bar n_x+l)$'th Landau level, i.e. $\frac{1}{2\epsilon} \simeq 
 (\bar n_x+\frac{1}{2}) +l\,$, 
$\bar n_x$ being a positive integer. One finds for $l_z$ the upper bound 
$l_z \leq \frac{n_x}{2n_x+1}$ (which gives back the classical condition 
for $n_x \rightarrow \infty$).
The fact that there are no bulk levels with positive angular momenta $l > n_x$ near the 
$n_x$'th Landau level is seen clearly on the numerically calculated spectrum 
as a 
function of the magnetic field, by following the levels from $B=0$ (see Fig. 
\ref{fig:disk} or 
\cite{nakamura}). 
Each time the band index $n_x$ is decreased by one, one level of positive $l$ 
is removed.

\item[ii.] 
$N_{\Phi} \gg 1$, {\em{edge states}}.\\ These  are associated with negative 
values of $l$, $|l| \geq N_\Phi$.
Using the results of section 2, we see that there are 
no levels if $\bar{\xi_0} > 1$, i.e. if $e^{-x_0} < 1-c$ (since $\sigma$ is an 
increasing function). This corresponds to the lower classical bound for the 
angular momentum
$l_z > (l_z)_{min}$, where $(l_z)_{min}= -N_{\Phi} \epsilon - \sqrt{2 N_{\Phi} \epsilon}$. 
For $\bar{\xi_0} \rightarrow 1^-$, the rescaled energy 
$\frac{1}{2\epsilon} = \frac{1}{2\bar{\epsilon}}$ becomes infinite.
The high energy part of the edge spectrum in a given band is thus
given by those $l_z=l\epsilon\,$ very close to (but higher than) 
$(l_z)_{min}$, i.e., by 
\begin{equation}
\frac{E}{\hbar \omega} \simeq\frac{1}{4 N_{\Phi}} ( |l| - N_{\Phi} )^2.
\label{high energies}
\end{equation}
In other words, the energy $E$ (for fixed $R$) varies like $(|l|-N_{\Phi})^2$ with the magnetic
field for such $|l| \gg N_{\Phi}$ (see Fig. \ref{fig:disk}). 
Note that $|l|$ may become infinite but this corresponds to 
$l_z \rightarrow 0$. Decreasing $|l|$, we go
towards lower energies. In the limit $N_{\Phi} \gg 1$, $N_{\Phi} \epsilon$ 
and $|l_z| = |l| \epsilon$  
become large, and to calculate the 
lower energy part of the edge spectrum we approximate 
$x_0\simeq \ln (\frac{|l|}{N_{\Phi}})$ and $a \simeq 
2 |l| \epsilon \simeq \frac{4}{c^2}$. Finally, by decreasing further $|l|$ below $N_{\Phi}$, 
the states transform gradually into Landau-like states located far apart 
from the boundary,
with energy levels closer to the Landau levels.

In the high field limit, so that edge states do correspond to large angular 
momenta 
with respect to $\hbar$, the semiclassical analysis is expected 
to work well. This is  expressed in equation (\ref{dwkb equation}) by the fact that the 
small parameter is $\frac{4\epsilon}{a} \sim \frac{2}{|l|} \sim \frac{2}{N_{\Phi} }$. 
For these states ${1}/{N_{\Phi}}$ may be thought as an effective Planck constant.
In Fig. \ref{fig:3} we show that the semiclassical bulk and edge energies are effectively very 
close to the exact energies for $N_{\Phi}=20$.

\item[iii.] $N_\Phi \ll 1$: {\em{low field limit}}. \\
When $\omega \rightarrow 0$, the exact
energies of the states of angular momentum $l\hbar$ tend to their zero-field values 
$E_0= \frac{\hbar^2 j_{l}^2}{2m R^2}$, where $j_{l}$ are the zeros of the Bessel function of order
$l$ (note that due to the time-reversal symmetry, states with opposite $l$ are degenerate in energy
at $B=0$, i.e. $j_{l}=j_{-l}$). For small fields and fixed $l$, 
$\epsilon = \frac{\hbar \omega}{2E}=O(\omega)$ and $l_z=l\epsilon$ is small. 
By
expanding the action $I_r(x_0)$ in (\ref{daction}) up to second order in the 
field we find
\begin{eqnarray} \label{low field action}
\frac{2 \pi}{\hbar |l|} I_r(x_0) & = & \nonumber 2\sqrt{z_0 -1} -2 \arccos({z_0}^{-\frac{1}{2}})
+ \frac{1-z_0^{-1}}{\sqrt{z_0-1}} (\delta z-\frac{2 N_{\Phi}}{l} ) \\
 & & \nonumber + (z_0-1)^{-\frac{3}{2}} \left( \, (-1 +\frac{3}{z_0} -\frac{2}{ z_0^2 } ) 
(\frac{\delta z^2}{4} - \frac{N_{\Phi} \delta z}{l} ) \right. \\
& & \left. +\frac{N_{\Phi}^2}{3l^2}
(-z_0 -3 + \frac{12}{z_0} -\frac{8}{ {z_0}^2} )\,\right) + O(\omega^i \delta z^j),  
\end{eqnarray}
where $z \equiv \frac{2 N_{\Phi} }{\epsilon \,l^2} = \frac{2 m R^2 E}{\hbar^2 l^2}$, 
$\lim_{\omega \rightarrow 0}z = z_0$, 
$\delta z \equiv  z-z_0$ and $i+j\geq 3$. Using (\ref{quantization edge}),
(\ref{integrated map equation}), (\ref{hatenergy}) and 
$\lim_{\omega \rightarrow 0} \epsilon I_r(x_0) = 0$, we obtain 
$\lim_{\omega\rightarrow 0} \bar{\xi}_0 = 1$. 
Therefore we can use
the `generalized EBK rules' developed in ii. of section 2 to quantize the action 
$I_r(x_0)$. Taking $\omega=0$ in (\ref{low field action}), we obtain that the square root 
energy $\sqrt{l^2 z_0}= \sqrt{\frac{2m R^2 E_0}{\hbar^2} }$ is equal to the first term of
the uniform approximation of the zeros of the Bessel function of order $l$, 
namely it satisfies (see for ex. \cite{abramowitz}):
\begin{equation} \label{zero field action}
\frac{\pi I_r(x_0)}{\hbar} = |l| \sqrt{z_0-1} - |l|
\arccos(z_0^{-\frac{1}{2}}) = \frac{2}{3} (-a_{n_x})^{\frac{3}{2}}.
\end{equation}
Our semiclassical
analysis gives thus the exact zero-field energies up to higher terms in 
this expansion (i.e.
with an error of the order of one percent for the lowest energies - including 
those with $l=0$).
This error decreases as a power law by increasing $|l|$ or $j_{l}$ 
\cite{abramowitz}. 
The EBK rules to quantized the classical action $I_r(x_0)$ 
(substituting $\pi (n_x+ \frac{3}{4})$ in the right-hand side of 
Eq. (\ref{zero field action})) give already very good approximations to
the exact energies \cite{keller} in the zero field case. 
We conclude that then the corrections due to proper treatment of  
the whispering gallery orbits are small.

For small but non-zero $\omega$, the action $I_r(x_0)$ remains
quantized through the same rules and is equal (in this approximation) 
to its zero field
value (\ref{zero field action}). It immediately follows from (\ref{low field action}) that
$\delta z= \frac{2 N_{\Phi}}{l} + O(\omega^2)$. Replacing this value into the action we 
obtain the energies up to second  order in $\omega$:
\begin{equation} \label{low field energies}
\frac{2mR^2 E}{\hbar^2}= j_{l}^2 + 2 l N_{\Phi} +\frac{N_{\Phi}^2}{3} 
( 1 + \frac{2 l^2}{j_{l}^2}) + O(\omega^3)
\end{equation}
This result agrees with recent semiclassical perturbative calculations at low fields 
\cite{gurevich}. However it does not coincide with the exact second-order 
perturbation theory result of Dingle \cite{dingle}, where a different factor
multiplying $N_\Phi^2$ in (\ref{low field energies}) was obtained (the two
expressions are equal in the limit of infinitely large energies).
This discrepancy should affect the magnetic susceptibility, which is
a measure of the field dependence of the levels, for low Fermi energies.
For sufficiently large energies (\ref{low field energies})
can be used to study the level crossings at
low $N_{\Phi}$'s. The values $N_{\Phi}$ at which there are level crossings 
are given by the solutions of 
the second degree equation:
\begin{equation} \label{level crossing}
 \frac{2 N_{\Phi}^2}{3} (\frac{l_1^2}{j_{l_1}^2}- \frac{l_2^2}{j_{l_2}^2}) +2 N_{\Phi} (l_1-l_2) 
+ j_{l_1}^2-j_{l_2}^2 = 0,
\end{equation}
where $j_{l_1}$ and $j_{l_2}$ are zeros of the Bessel functions of order 
respectively $l_1$ and $l_2$. 

\end{itemize} 

For arbitrary $N_{\Phi}$ and $l$ we can
compute the WKB wavefunction using the results of the previous section 
and the
formula (\ref{integrated map equation}) and (\ref{approximate sigma}).
 It is given by ($\eta$ is a small parameter):
\begin{eqnarray} \label{dwkb function}
f_{WKB} (x) & = & \nonumber  Q(x)^{-\frac{1}{4}}\, 
e^{\frac{i}{\hbar} \pi I_r(x) },\,\,\,\,\,\,{\rm{if}}\,\, -e^{-x} +1-c \gg \eta  \\
        & = & \nonumber \frac{2 \sqrt{\pi}}{(c-c^2)^{\frac{1}{6}} } ( \frac{a}{8 
        \epsilon })^{\frac{1}{6}} {\rm{Ai}} \left(
        (\frac{a\sqrt{2c}}{4 \epsilon (1-c) })^{\frac{2}{3}} ( -e^{-x}+1-c)\right),
        \,\,\,\,\,\,{\rm{if}}\,\, |e^{-x}-1+c| \ll c  \\
        & = & \nonumber 2 (-Q(x))^{-\frac{1}{4}}\,
        \sin \left( \frac{\pi I_r(x)}{\hbar}+\frac{\pi}{4} \right), \,\,\,\,\,\,{\rm{if}} \,\,
        e^{-x} -1+c \gg \eta, \,\, -e^{-x} +1+c \gg \eta \\
        & = & \nonumber \frac{2 \sqrt{\pi}}{(c+c^2)^{\frac{1}{6}} } ( \frac{a}{8 
        \epsilon })^{\frac{1}{6}} \, \left\{  \, \sin ( \frac{\pi}{2\bar{\epsilon}} ) \,
        {\rm{Ai}} \left( \,  
        (\frac{a \sqrt{2c}}{4 \epsilon (1+c)})^{\frac{2}{3}} ( e^{-x}-1-c ) \right) + \right.\\
        & &\,\,\,\,\,\,\,\,\, \nonumber + \left. \cos (\frac{\pi}{2\bar{\epsilon}}) \, 
        {\rm{Bi}}  \left( \, 
        ( \frac{a\sqrt{2c}}{4 \epsilon (1+c)})^{\frac{2}{3}} ( e^{-x}-1-c ) \right) \,\right\}
        \,\,\, \,\,\,{\rm{if}}\,\, |e^{-x}-1-c| \ll c  \\
         & = & Q(x)^{-\frac{1}{4}}\,
        \left\{ \sin(\frac{\pi}{2\bar{\epsilon}}) \,e^{-\frac{i}{\hbar} \pi I_r(x) }
        + 2\cos(\frac{\pi}{2\bar{\epsilon}}) \,e^{\frac{i}{\hbar} \pi I_r(x)}  \right\},
        \,\,\,\,\,\, {\rm{if}}\,\, e^{-x}-1-c \gg \eta.
\end{eqnarray}


Suppose that one introduce an Aharonov-Bohm flux line through the origin of the disc, and let us
study the transport of charges induced by varying the flux $\Phi$. The
vector potential is changed by ${\bf{A}} \rightarrow {\bf{A}}_0 + \frac{\Phi}{2 \pi r} 
{\bf{e}}_{\theta}$, where ${\bf{A}}_0$ is the vector potential created by the uniform magnetic 
field. The Schr\"odinger equation for this system is still of the form 
(\ref{dsturm liouville})
but with $l$ replaced by $l + \alpha$, where $\alpha = {\Phi}/{\Phi_0}$, therefore all our 
results still apply modulo this substitution. The spectrum is identical for 
$\alpha$'s which 
differ by integer values (gauge invariance). Changing 
continuously the flux $\Phi$ is equivalent to an electromotive force 
(e.m.f.) through 
the system. Consider $\alpha$ varying continuously from $0$ to $1$. Then all eigenstates and
energies of the system evolve adiabatically so that edge levels lower their energies and
fall (at $\alpha =1$) into the level situated just below them. 
Bulk levels are nearly not affected except those
with positive $l$, whose energies $E = \hbar \omega (\bar n_x + 
l +\alpha + \frac{1}{2})$ is increased by 
$\hbar \omega$, the electron moving from one Landau level into another (see 
Fig. \ref{fig:3}). 
Introducing the Fermi energy $E_F$ for an independent electron gas at
zero temperature, we see that the effect of changing the flux by one unit 
is to remove $n_x$ electrons from the center of the disc and to transfer them 
to the
boundary if $E_F$ is between the $n_x$'th and the $(n_x+1)$'th Landau level
(as in Laughlin's argument \cite{QHE}).
This conclusion is not specific to the disc geometry, since it is due to
positive $l$ states which are almost insensitive to  
the boundary when $N_\Phi>>1$.
 This scenario is therefore generalizable to arbitrary billiards 
in the strong magnetic fields.

\section{Conclusion}
We have presented a calculation of the energy spectrum of an electron 
in a perpendicular and uniform magnetic field for the half plane and the disc 
geometry. These situations are the simplest to describe features due to the 
presence of both perpendicular field and a boundary. 
In fact, explicit calculations can be made in both cases using the existence 
of a constant of
motion resulting from the symmetry of the system (namely the
momentum in the direction parallel to the boundary for the half plane, and 
the angular momentum for the disc). One can then reduce these two-dimensional 
problems into one-dimensional ones and use semiclassical methods to solve them. 

Because of the inherent splitting between bulk and edge classical trajectories,
in the 
standard EBK approach one still have a clear separation between bulk and edge
states. A more correct semiclassical approach removes this splitting by taking
into account the 
coalescence of the caustics with the boundary which are due to edge orbits 
making 
nearly full circles before being reflected or to bulk orbits
very close to the boundary. This removes the EBK gap between 
bulk and edge energies and results in a semiclassical spectrum which 
approximates surprisingly well the lowest levels, 
as far as we checked from our numerical calculations.
Blaschke and Brack
did independently reach similar conclusions using periodic orbits techniques
in their recent work on the disc billiard \cite{blaschke}.
In the quantum system under usual boundary conditions 
(like the Dirichlet case here considered), the notion of edge and bulk states 
is thus not anymore well-defined. To restore a precise meaning 
of such a dichotomy in the quantum case and describe for instance the 
transition from edge to bulk states in the context of the integer quantum Hall
 effect, 
one has to consider more general boundary conditions \cite{we} pertaining to 
the family defined by Atiyah, 
Patodi and Singer \cite{aps}. The extension of our results to general 
(non-integrable) billiards
with smooth boundaries could be made or by using the more elaborate 
path-integration
semiclassical methods, or, for almost circular billiards, by expanding the 
wavefunction on the
WKB disc wavefunction. Such billiards are interesting because of the
existence in zero field of regimes of localization of the wavefunction in the 
angular momentum space
\cite{frahm}. Our approach of deriving the uniform approximation 
for the separable systems can be of relevance in connection
with the recent work on uniform approximation 
of the diffraction contribution to the semiclassical spectral density
for general magnetic field free billiards \cite{uzy}. 

{\bf Acknowledgement}

This work was supported in part by a grant from the Israel Science
foundation and by the fund for promotion of the research at the Technion.
R.~N. acknowledges support by BSF grant 01-4-32842.

\renewcommand{\thesection}{\arabic{section}}
\setcounter{section}{2}
\renewcommand{\thefigure}{\thesection\alph{figure}}
\setcounter{figure}{0}

\begin{figure}
\begin{center}
  	\leavevmode
	\epsfig{file=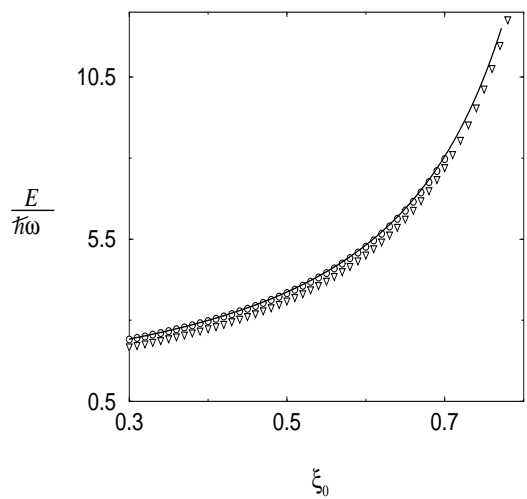,height=7cm,width=7cm,angle=270}
\end{center}
\caption{The energy spectrum for $n_x=0$: the full line is the exact spectrum, triangles 
are
the zeros of the second branch and circles - the zeros of the third branch. }
\label{fig:2a}
\end{figure}

\newpage

\begin{figure}
\begin{center}
  	\leavevmode
	\epsfig{file=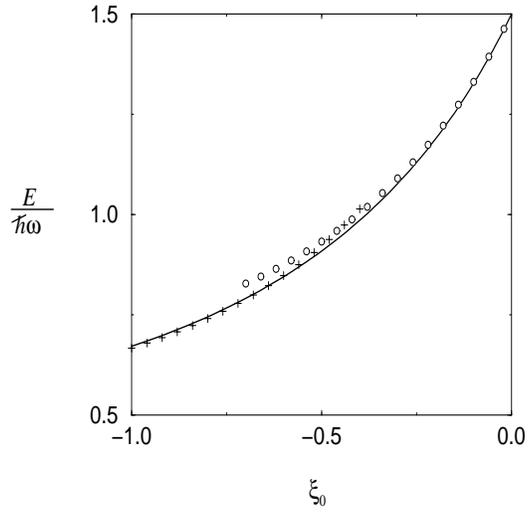,height=7cm,width=7cm,angle=270}
\end{center}
\caption{The energy spectrum for $n_x = 0$: circles are
the zeros of the fourth branch, otherwise notations like in Fig. 2a. }
\label{fig:2b}
\end{figure}

\begin{figure}
\begin{center}
  	\leavevmode
	\epsfig{file=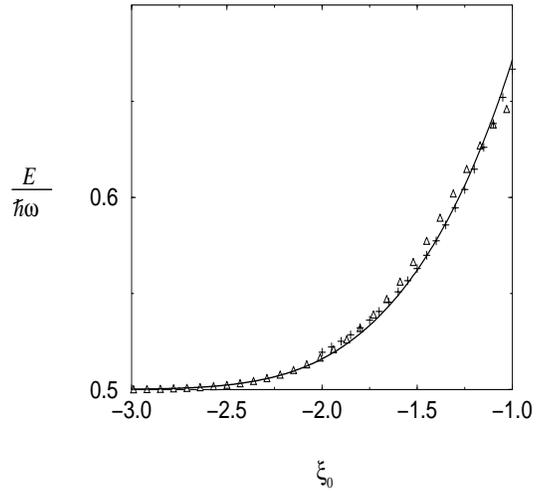,height=7cm,width=7cm,angle=270}
\end{center}
\caption{The energy spectrum for $n_x=0$: triangles are
the zeros of the fifth branch, otherwise notations like in Fig. 2b. }
\label{fig:2c}
\end{figure}

\newpage

\setcounter{figure}{2}
\renewcommand{\thefigure}{\arabic{figure}}

\begin{figure}
\begin{center}
 	\leavevmode
	\epsfig{file=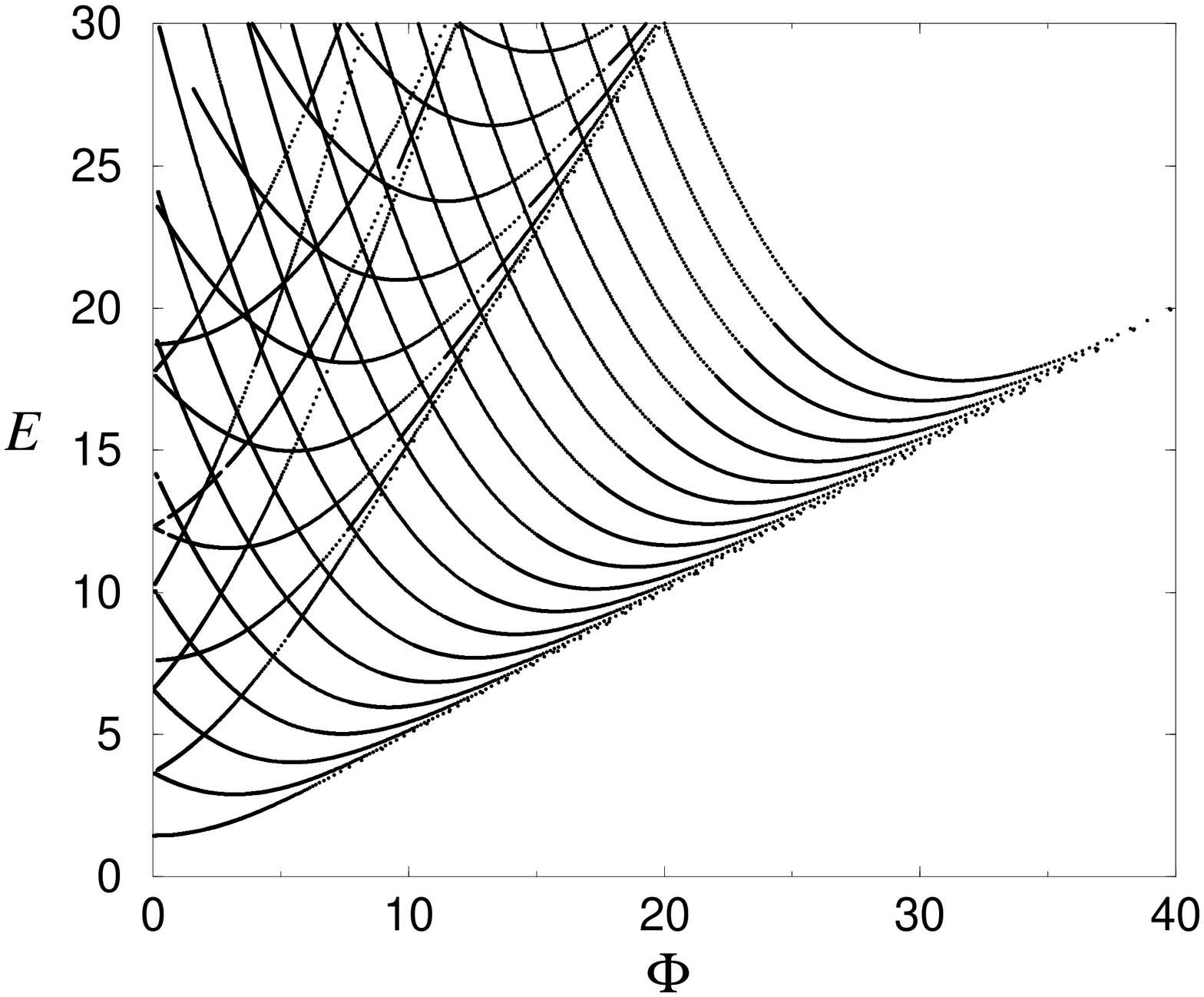,height=7cm,width=7cm,angle=270}
\end{center}
\caption{The energy spectrum of the disc as function of the magnetic flux. 
The energy $E$ in the
vertical axis is in units of $\frac{2\hbar^2}{mR^2}$.}
\label{fig:disk}
\end{figure}

\begin{figure}
\begin{center}
 	\leavevmode
	\epsfig{file=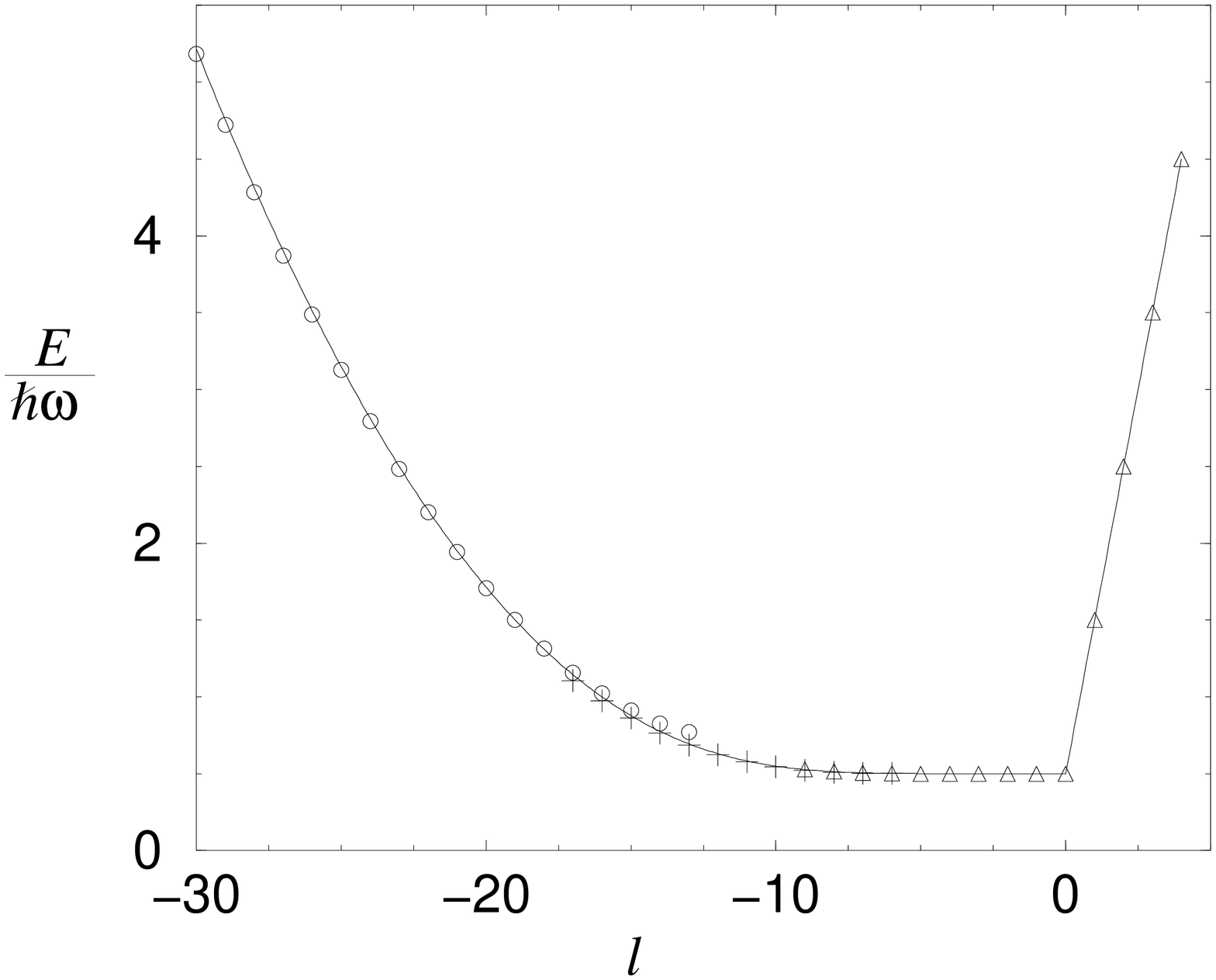,height=7cm,width=7cm,angle=270}
\end{center}
\caption{The energy spectrum of the disc as function of the angular
momentum ($N_\Phi=20$): circles are the zeros of the third branch,
pluses - the zeros of the fourth branch and triangles are the zeros of the 
fifth branch.}
\label{fig:3}
\end{figure}

\end{document}